\documentclass[twocolumn,aps,prc,superscriptaddress,nofootinbib,showkeys,floatfix,singlecolumn]{revtex4}

\usepackage[english]{babel}

\usepackage[letterpaper,top=2cm,bottom=2cm,left=2cm,right=2cm,marginparwidth=1.7cm]{geometry}

\usepackage{amsmath}
\usepackage{graphicx}
\usepackage[colorlinks=true, allcolors=blue]{hyperref}
\allowdisplaybreaks

\begin{document}
\title{Exploring system size dependence of jet modification in heavy-ion collisions}
\author{Yang He}\affiliation{Institute of Frontier and Interdisciplinary Science, Shandong University, Qingdao, 266237, China}\affiliation{Key Laboratory of Particle Physics and Particle Irradiation, Ministry of Education, Shandong University, Qingdao, Shandong, 266237, China}
\author{Mengxue Zhang}\affiliation{Institute of Frontier and Interdisciplinary Science, Shandong University, Qingdao, 266237, China}\affiliation{Key Laboratory of Particle Physics and Particle Irradiation, Ministry of Education, Shandong University, Qingdao, Shandong, 266237, China}
\author{Maowu Nie}\email{maowu.nie@sdu.edu.cn}\affiliation{Institute of Frontier and Interdisciplinary Science, Shandong University, Qingdao, 266237, China}\affiliation{Key Laboratory of Particle Physics and Particle Irradiation, Ministry of Education, Shandong University, Qingdao, Shandong, 266237, China}
\author{Shanshan Cao}\email{shanshan.cao@sdu.edu.cn}\affiliation{Institute of Frontier and Interdisciplinary Science, Shandong University, Qingdao, 266237, China}\affiliation{Key Laboratory of Particle Physics and Particle Irradiation, Ministry of Education, Shandong University, Qingdao, Shandong, 266237, China}
\author{Li Yi}\email{li.yi@sdu.edu.cn}\affiliation{Institute of Frontier and Interdisciplinary Science, Shandong University, Qingdao, 266237, China}\affiliation{Key Laboratory of Particle Physics and Particle Irradiation, Ministry of Education, Shandong University, Qingdao, Shandong, 266237, China}	

\begin{abstract}

In relativistic heavy-ion collisions, jet quenching in quark-gluon plasma (QGP) has been extensively studied, revealing important insights into the properties of the color deconfined nuclear matter. Over the past decade, there has been a surge of interest in the exploration of QGP droplets in small collision systems like $p$+$p$ or $p$+A collisions driven by the observation of collective flow phenomena. However, the absence of jet quenching, a key QGP signature, in these systems poses a puzzle. Understanding how jet quenching evolves with system size is crucial for uncovering the underlying physics. In this study, we employ the linear Boltzmann transport (LBT) model to investigate jet modification in $^{96}$Ru+$^{96}$Ru, $^{96}$Zr+$^{96}$Zr, and $^{197}$Au+$^{197}$Au collisions at $\sqrt{s_\mathrm{NN}}=200$~GeV. Our findings highlight the system size sensitivity exhibited by jet nuclear modification factor ($R_\mathrm{AA}$) and jet shape ($\rho$), contrasting to the relatively weak responses of jet mass ($M$), girth ($g$) and momentum dispersion ($p_\mathrm{T}{D}$) to system size variations. These results offer invaluable insights into the system size dependence of the QGP properties and await experimental validation at the Relativistic Heavy-Ion Collider.

\end{abstract}

\keywords{relativistic heavy-ion collisions, quark-gluon plasma, jet quenching, jet substructure, system size dependence}

\maketitle

\section{Introduction}

Extensive studies from the Relativistic Heavy-Ion Collider (RHIC) and the Large Hadron Collider (LHC) have demonstrated that a nearly perfect fluid of strongly interacting quark-gluon plasma (QGP) is produced in these energetic heavy-ion collisions~\cite{Gyulassy:2004zy,Jacobs:2004qv, Busza:2018rrf}. Considering the created QGP only lives for a very short time, a wide range of indirect observables based on detectable particles have been proposed~\cite{Harris:2024aov}. In particular, anisotropic collective flow and jet quenching have been found to be key evidence for the existence of QGP in high-energy large-nuclei collisions over the past few decades. Azimuthal anisotropic flow serves as an important tool to reveal collectivity of the QGP, which is related to the transport properties, e.g. specific viscosity, of the hot medium that determine the amount of azimuthal asymmetry transferred from the initial geometric distribution of the QGP into the final momentum space distribution of the produced particles~\cite{Ollitrault:1992bk,Bernhard:2019bmu}. Jet quenching, resulting from energy loss of high-energy partons through the hot medium, reveals the opacity of the medium to these partons produced from the initial hard collisions~\cite{Wang:1992qdg,Qin:2015srf,Majumder:2010qh}. The large jet transport coefficient $\hat{q}$ extracted from the jet quenching data at RHIC and LHC, which is an order of magnitude higher than that inside a cold nucleus~\cite{JET:2013cls,JETSCAPE:2021ehl}, provides additional evidence of the quark-gluon degrees of freedom of the hot medium created in relativistic heavy-ion collisions. 

The observations of collective flow in small systems, such as those produced by proton-proton ($p$+$p$) and proton-nucleus ($p$+A) collisions, have sparked lively discussions on the existence of QGP in these small systems~\cite{CMS:2010ifv,ATLAS:2012cix, ALICE:2012eyl,PHENIX:2013ktj,STAR:2015kak,STAR:2022pfn,Noronha:2024dtq}. However, jet quenching has not been observed in such systems so far~\cite{ATLAS:2022iyq,PHENIX:2015fgy,ALICE:2017svf,Yi:2019czm}. Therefore, the nature of the medium produced in small systems is still to be revealed. A system size scan of jet quenching can help bridge the gap between parton energy loss in large and small collision systems, and search for the lower limit of the medium size where signals of jet quenching still remain. Although one can obtain QGP with various sizes by selecting different centrality classes of heavy-ion collisions, it was found in Ref.~\cite{Loizides:2017sqq} that significant event selection and geometry biases may be introduced to jet quenching in peripheral collisions, which cause suppression of jet spectra even in the absence of parton energy loss. For this reason, collisions using different types of nuclei have been proposed at the LHC energy~\cite{Citron:2018lsq,Huss:2020dwe} for obtaining QGP media with different sizes but similar geometry, and considerable theoretical efforts~\cite{Zigic:2018ovr,Shi:2018izg,Katz:2019qwv,Huss:2020whe,Liu:2021izt,Li:2021xbd} have also been devoted to understanding parton-medium interactions in these different collision systems. While the nuclear modification of energetic hadrons have been extensively studied in these pioneer work, dependence of fully reconstructed jet observables on colliding nuclei has rarely been discussed yet. Recently, measurements on full jets in $^{96}$Zr+$^{96}$Zr and $^{96}$Ru+$^{96}$Ru collisions conducted at RHIC~\cite{He:2024roj} provide an invaluable opportunity to investigate interactions between high-energy partons and the QGP with a size between $p$+A and Au+Au collisions. The purpose of the present work is to explore the sensitivity of different jet observables to the size of a color deconfined medium, which may provide a timely theoretical reference to the ongoing jet measurements at RHIC above.

Ever since jet quenching has been confirmed experimentally, a variety of jet observables have been utilized to infer the properties of jet-QGP interactions~\cite{Blaizot:2015lma,Kogler:2018hem,Andrews:2018jcm,Connors:2017ptx,Cao:2020wlm,Cao:2022odi,Cao:2024pxc}. For example, the nuclear modification factor $R_\mathrm{AA}$ (or $R_\mathrm{CP}$) quantifies the variation of jet spectrum in nucleus-nucleus (A+A) collisions with respect to its proton-proton ($p$+$p$) reference (or reference from peripheral A+A collisions)~\cite{Wang:2004dn,PHENIX:2001hpc,STAR:2017ieb}. Jet mass characterizes the virtual scale of the initial hard parton that generates the jet and medium modification of this scale during parton showers~\cite{Majumder:2014gda,ALICE:2017nij,Park:2018acg, Casalderrey-Solana:2019ubu}. Jet shape $\rho$ and jet girth $g$, constructed to explore intra-jet structure, provide information on the medium modification of energy distribution of particles within a jet~\cite{CMS:2013lhm, Cunqueiro:2015dmx, ATLAS:2014dtd, CMS:2012nro, Yan:2020zrz, Tachibana:2017syd,JETSCAPE:2023hqn,Tang:2023epb}. Jet momentum dispersion $p_\mathrm{T}D$ is developed to measure the hardness of jet fragmentation, which may be used to distinguish between quark and gluon jets, and infer the underlying dynamics of parton showers and hadronization processes~\cite{ALICE:2018dxf,Chen:2022kic}. 
These tools allow for a multi-dimensional investigation on the interactions between energetic jet partons and a dense nuclear medium.


In this work, we study the system size dependence of jet-QGP interactions by comparing nuclear modification of jets between Zr+Zr (Ru+Ru) and Au+Au collisions at the top RHIC energy. The initial jets from hard scatterings are generated by PYTHIA~8 simulations~\cite{Sjostrand:2014zea,Sjostrand:2006za}, with formation time of each jet parton added according to the parton shower history~\cite{Zhang:2022ctd}. The evolution of the QGP is modeled with a (3+1)-dimensional viscous hydrodynamic model CLVisc~\cite{Pang:2018zzo,Wu:2018cpc,Wu:2021fjf}. The interactions between high-energy jet partons and the QGP are described by a linear Boltzmann transport (LBT) model~\cite{Luo:2023nsi} that accounts for both elastic and inelastic scatterings of jet partons inside a color deconfined nuclear matter. Comparing to earlier studies on impacts of system size on jet energy loss, we investigate not only the nuclear modification factor of jets, but also their various substructure observables, including jet shape, jet mass, jet girth, and jet momentum dispersion. Specifically, we explore the sensitivities of these observables to the size of QGP, which may help future jet measurements search for signals of QGP inside a small collision system.


The remainder of this paper is organized as follows. In Sec.~\ref{sec:model}, we describe the numerical framework of jet-QGP interactions we use for calculating the jet observables in relativistic heavy-ion collisions. In Sec.~\ref{sec:results}, we investigate how the jet yield and various intra-jet observables depend on the size of colliding nuclei at the top RHIC energy. In the end, we summarize in Sec.~\ref{sec:summary}.

\section{Jet transport inside the QGP}
\label{sec:model}

We use the linear Boltzmann transport (LBT) model~\cite{Cao:2016gvr,Luo:2023nsi} to simulate interactions between jet partons and the QGP. In LBT, the phase space distribution of jet partons $f_a(\vec{x}_a,\vec{p}_a,t)$ evolves according to the Boltzmann equation as follows:
\begin{equation} 
\label{eq:bolz}
\large
p_a \cdot \partial f_a = E_a\left[C^{\rm el}(f_a) + C^{\rm inel}(f_a)\right],
\end{equation}
where $p_a=(E_a,\vec{p}_a)$ denotes the four momentum of a jet parton with the on-shell condition satisfied. Zero mass is assumed for thermal partons inside the QGP. On the right hand side, $C^{\rm el}$ and $C^{\rm inel}$ are collision integrals for elastic and inelastic scatterings respectively. If one only considers medium modification on the jet parton distribution, but not the inverse process, the collision integrals can be simplified to linear functions of $f_a$.

From the linearized equation with respect to $f_a$, one may extract the scattering rate of a single parton at a given momentum state. For instance, inside an isotropic static medium at temperature $T$, the elastic scattering rate of a parton with energy $E_a$ is given by
\begin{align}
\label{eq:rate}
\large
\Gamma_a^\mathrm{el}&(E_a,T)=\sum_{b,(cd)}\frac{\gamma_b}{2E_a}\int \prod_{i=b,c,d}\frac{d^3p_i}{E_i(2\pi)^3} f_b(E_b,T) \nonumber\\
&\times [1\pm f_c(E_c,T)][1\pm f_d(E_d,T)] S_2(\hat{s},\hat{t},\hat{u})\nonumber\\
&\times (2\pi)^4\delta^{(4)}(p_a+p_b-p_c-p_d)|\mathcal{M}_{ab\rightarrow cd}|^2,
\end{align}
where the sum runs over all possible $2\rightarrow 2$ scattering channels, $|\mathcal{M}_{ab\rightarrow cd}|^2$ represents the scattering amplitude square of a specific channel, which is proportional to the square of the strong coupling strength ($\alpha_\mathrm{s}^2$). Here, $b$ denotes an incoming thermal parton inside the medium, while $c$ and $d$ are the outgoing partons from the scattering. Bose and Fermi distributions -- $f_i$ ($i=b,c,d$) -- are taken for thermal gluons and quarks, respectively. The factor $\gamma_b$ represents the spin-color degrees of freedom for $b$. To avoid divergence in the leading-order matrix element applied in this work, a double-theta function $S_2(\hat{s},\hat{t},\hat{u})=\theta(\hat{s}\ge 2\mu_\mathrm{D}^2)\,\theta(-\hat{s}+\mu^2_\mathrm{D}\le \hat{t} \le -\mu_\mathrm{D}^2)$ is implemented, where $\hat{s}$, $\hat{t}$, $\hat{u}$ are the Mandelstam variables and $\mu_{\rm{D}}^2=4\pi\alpha_\mathrm{s}T^2(N_c + N_f/2)/3$ is the Debye screening mass, with $N_c$ and $N_f$ the color and flavor numbers, respectively. 
 
The inelastic scattering rate is related to the number of medium-induced gluons per unit time as 
\begin{equation}
\label{eq:gluonnumber}
\large
\Gamma_a^\mathrm{inel} (E_a,T,t) = \int dzdk_\perp^2 \frac{1}{1+\delta^{ag}}\frac{dN_g^a}{dz dk_\perp^2 dt},
\end{equation}
where the spectrum of gluon emission -- ${dN_g^a}/(dz dk_\perp^2 dt)$ -- is taken from the higher-twist energy loss calculation~\cite{Wang:2001ifa,Zhang:2003wk,Majumder:2009ge}, with $z$ and $k_\perp$ the fractional energy and transverse momentum of the emitted gluon with respect to its parent parton $a$. The $\delta^{ag}$ function is applied in the above equation to avoid double counting in evaluating the $g\rightarrow gg$ rate from its splitting function. The medium-induced gluon spectrum here is proportional to the jet transport coefficient $\hat{q}$, which characterizes the transverse momentum broadening square of a jet parton per unit time due to elastic scatterings, and thus can be evaluated using Eq.~(\ref{eq:rate}) with a weight factor $[\vec{p}_c-(\vec{p}_c\cdot\hat{p}_a)\hat{p}_a]^2$ added inside the integral. In the end, the only free parameter in LBT is $\alpha_\mathrm{s}$. In this work, we set $\alpha_\mathrm{s}=0.3$, which can provide a reasonable description of the inclusive jet yield suppression in Au+Au collisions at $\sqrt{s_\mathrm{NN}}=200$~GeV~\cite{Zhang:2022ctd}.


In this work, we use PYTHIA~\cite{Sjostrand:2014zea,Sjostrand:2006za} to generate vacuum jets as input for the LBT model. The formation time of each parton inside a vacuum jet is evaluated according to 
\begin{equation} 
\large
\tau_\mathrm{form}=\frac{2Ez(1-z)}{k_\perp^2},
\end{equation} 
where $E$ is the energy of the parton's ancestor directly produced from an initial hard scattering in PYTHIA, $z$ is energy fraction taken by the parton from its ancestor, and $k_\perp$ is the transverse momentum of the parton relative to its ancestor. Partons emanating from a given PYTHIA event are assumed to originate from the same location, randomly drawn from the binary collision vertices given by the Monte-Carlo Glauber model for nucleus-nucleus collisions. Each of these partons then streams freely in space according to its velocity before it enters LBT and starts interacting with the QGP at $\tau_\mathrm{init} = \max(\tau_\mathrm{form}, \tau_0)$, where  $\tau_0=0.6$~fm/$c$ is the commencement time of the hydrodynamic evolution of the QGP. The possible medium modification of a jet parton before it reaches the formation time, or at the high virtuality stage, is ignored in our current calculation, which can be taken into account by the MATTER model~\cite{Cao:2017qpx} within the JETSCAPE framework~\cite{JETSCAPE:2017eso}. 

The QGP medium is simulated using the (3+1)-dimensional viscous hydrodynamic model CLVisc~\cite{Pang:2018zzo,Wu:2018cpc,Wu:2021fjf}. The hydrodynamic parameters are tuned based on the soft hadron spectra observed in Au+Au collisions at $\sqrt{s_\mathrm{NN}}=200$~GeV, and applied to Zr+Zr and Ru+Ru systems at the same beam energy. These hydrodynamic simulations provide the spacetime evolution profiles of the temperature and fluid velocity fields of the QGP, based on which we boost each parton into the local rest frame of the medium at each time step and update its momentum based on its elastic and inelastic scattering rates, Eqs.~(\ref{eq:rate}) and~(\ref{eq:gluonnumber}), inside the medium. We iterate these interactions for each time step until a parton under consideration exits the QGP boundary, the hypersurface of $T_\mathrm{pc}=165$~MeV in this work.

In the LBT model, we track the evolution of not only the jet partons fed from PYTHIA and their emitted gluons, but also thermal partons scattered out of the medium background (named as ``recoil" partons) by these jet partons and the associate energy-momentum depletion left inside the medium (modeled as ``negative" partons). We call jet partons together with their medium-induced gluons ``medium-modified jet shower partons", while recoil and negative partons ``jet-induced medium excitation" or ``medium response". Experimentally, one cannot accurately identify the origin of each constituent inside a jet, and therefore jets reconstructed from heavy-ion collision events inevitably contain contributions from medium response. In this work, we will include both recoil and negative partons in our jet finding algorithm, with the energy-momentum of the latter subtracted from all jet observables. This procedure is necessary to guarantee energy-momentum conservation during jet-medium interactions, and has been shown to be crucial for a quantitative understanding of the suppression, flow, and substructure of jets in earlier LBT calculations~\cite{He:2018xjv,He:2022evt,Luo:2018pto}.

\section{Jet modification in Au+Au, Zr+Zr and Ru+Ru collisions}
\label{sec:results}

In this section, we present numerical results of inclusive jet suppression and a series of intra-jet observables in Au+Au, Ru+Ru and Zr+Zr collisions at $\sqrt{s_\mathrm{NN}}=200$~GeV. Medium-modified jets in heavy-ion collisions are generated using the LBT model with PYTHIA events as input, while results directly from the PYTHIA simulation are also presented as a baseline of $p$+$p$ collisions. Due to the lack of a solid hadronization scheme, we discuss jet observables at the partonic level in this work. This prevents us from studying observables that are sensitive to the energies of individual hadrons, e.g. the jet fragmentation function, but may have less impact on observables that mainly rely on the total energy of constituents within a jet cone. We expect our qualitative conclusions below on the sensitivities of various observables to the system size of nuclear collisions still hold at the hadronic level.

\subsection{Nuclear modification factor}

The suppression of jet yields in A+A collisions compared to $p$+$p$ collisions can be qualified using the nuclear modification factor $R_\mathrm{AA}$. The magnitude of $R_\mathrm{AA}$ is driven by the interplay between the slope of jet spectra and the amount of jet energy loss, i.e., energy flowing outside the jet cone. Both elastic and inelastic scatterings can cause jet energy loss,   
which are correlated with the path length of jets inside the QGP. Therefore, a comparison of $R_\mathrm{AA}$ across QGP systems of different sizes can improve our understanding on the path length dependence of the jet energy loss through the QGP. 

We use the Fastjet package for jet reconstruction with the anti-$k_\mathrm{T}$ algorithm~\cite{Cacciari:2005hq,Cacciari:2011ma}. Partons for jet reconstruction are required to be within the mid-rapidity range $|\eta| < 1$ and have a transverse momentum $p_\mathrm{T} > 0.2$~GeV/$c$. Each individual jet is constrained to lie within the rapidity range of $|\eta_{\text{jet}}| < 1 - R$, where $R$ is the jet cone size. A jet is rejected if its area $A_{\text{jet}}$ is below certain thresholds: $A_{\text{jet}} < 0.07$ for $R = 0.2$, $A_{\text{jet}} < 0.2$ for $R = 0.3$, and $A_{\text{jet}} < 0.4$ for $R = 0.4$. These area cuts follow experimental practices to reduce the contribution of uncorrelated background jets~\cite{STAR:2017hhs,STAR:2020xiv}. The same jet sample is utilized for both jet shape $\rho$ and jet mass $M$ calculations in the next two subsections to ensure consistency with the experimental setups.

\begin{figure}[tbp!]
	\centering
    \vspace{-20pt}
	\includegraphics[scale=0.35]{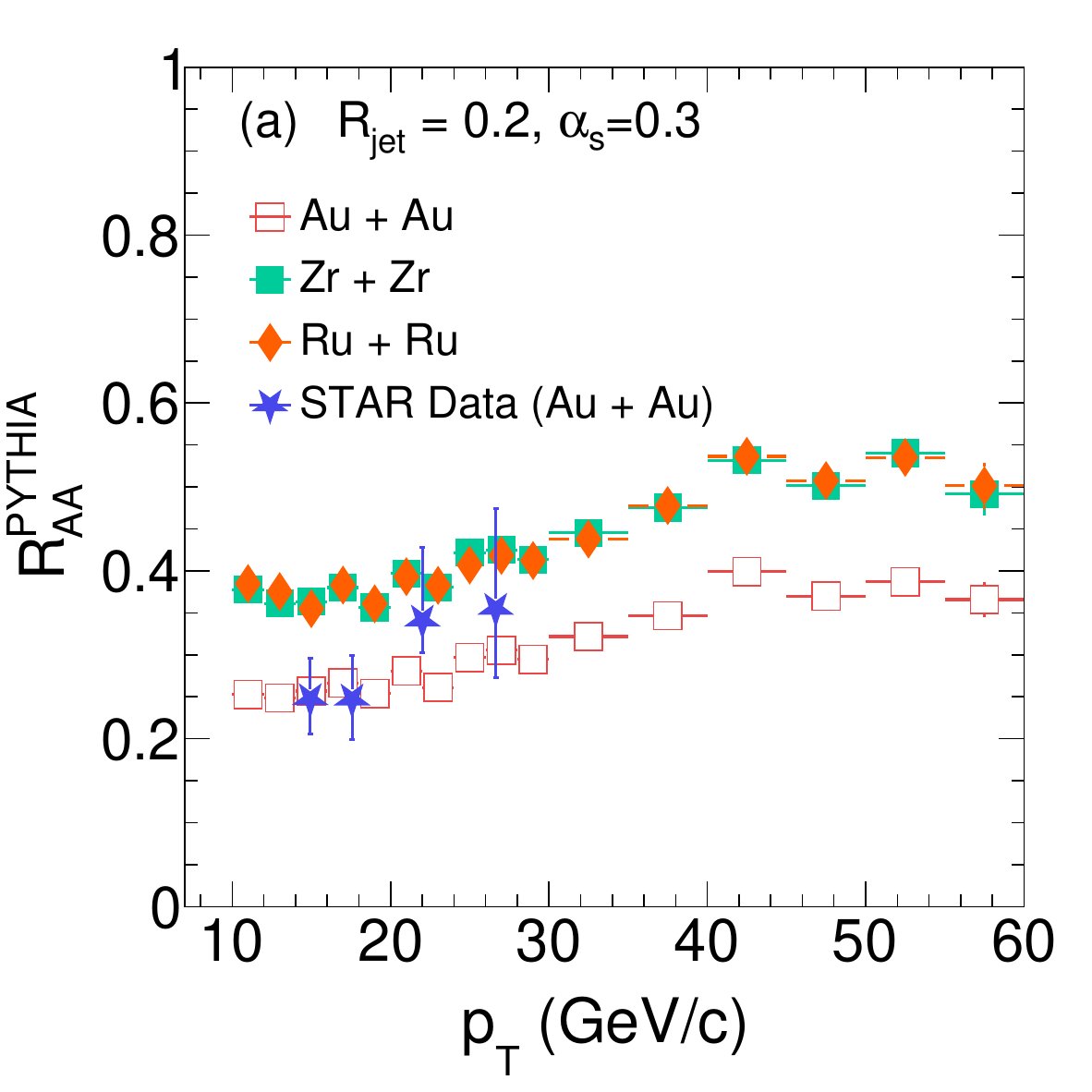}\\
	\includegraphics[scale=0.35]{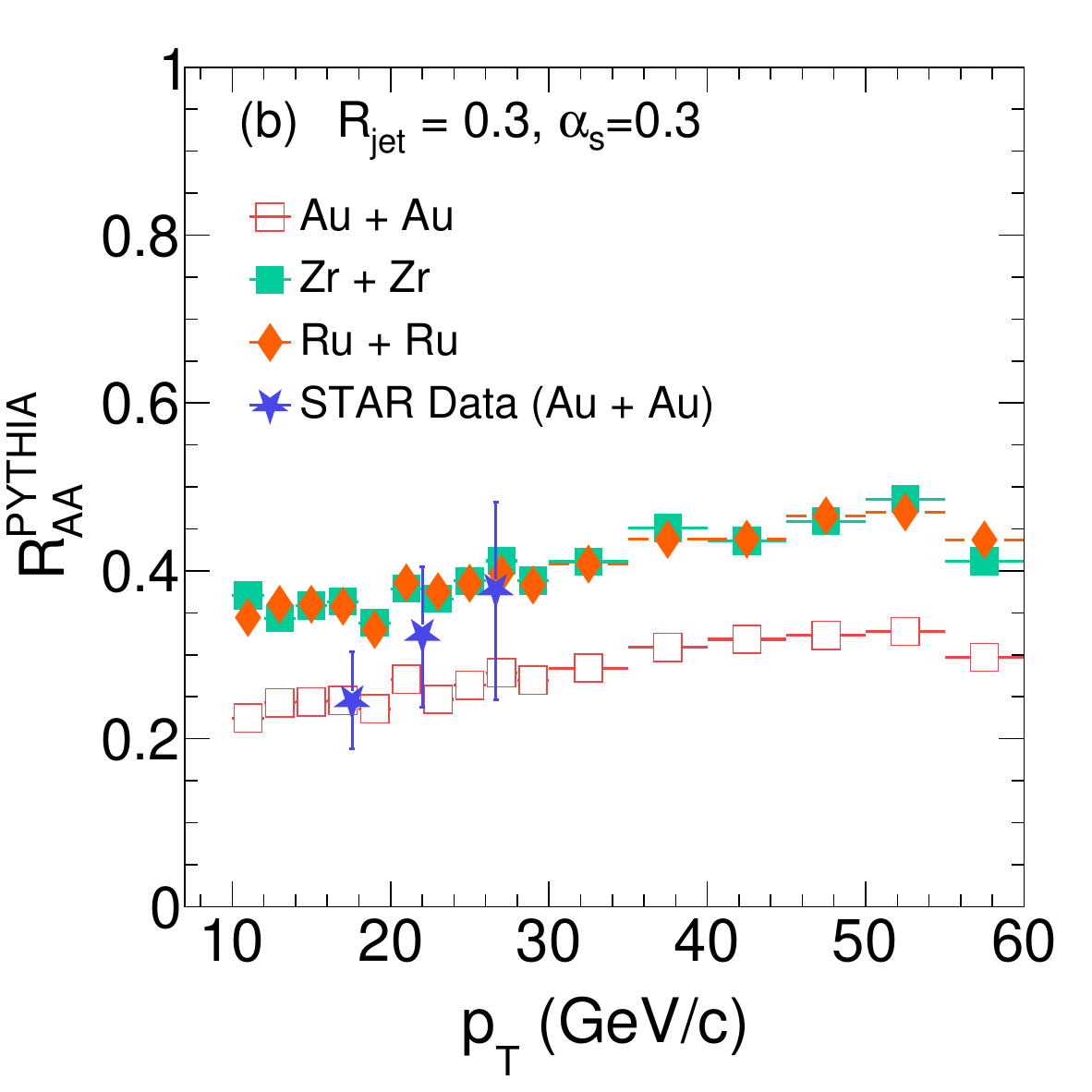}\\
	\includegraphics[scale=0.35]{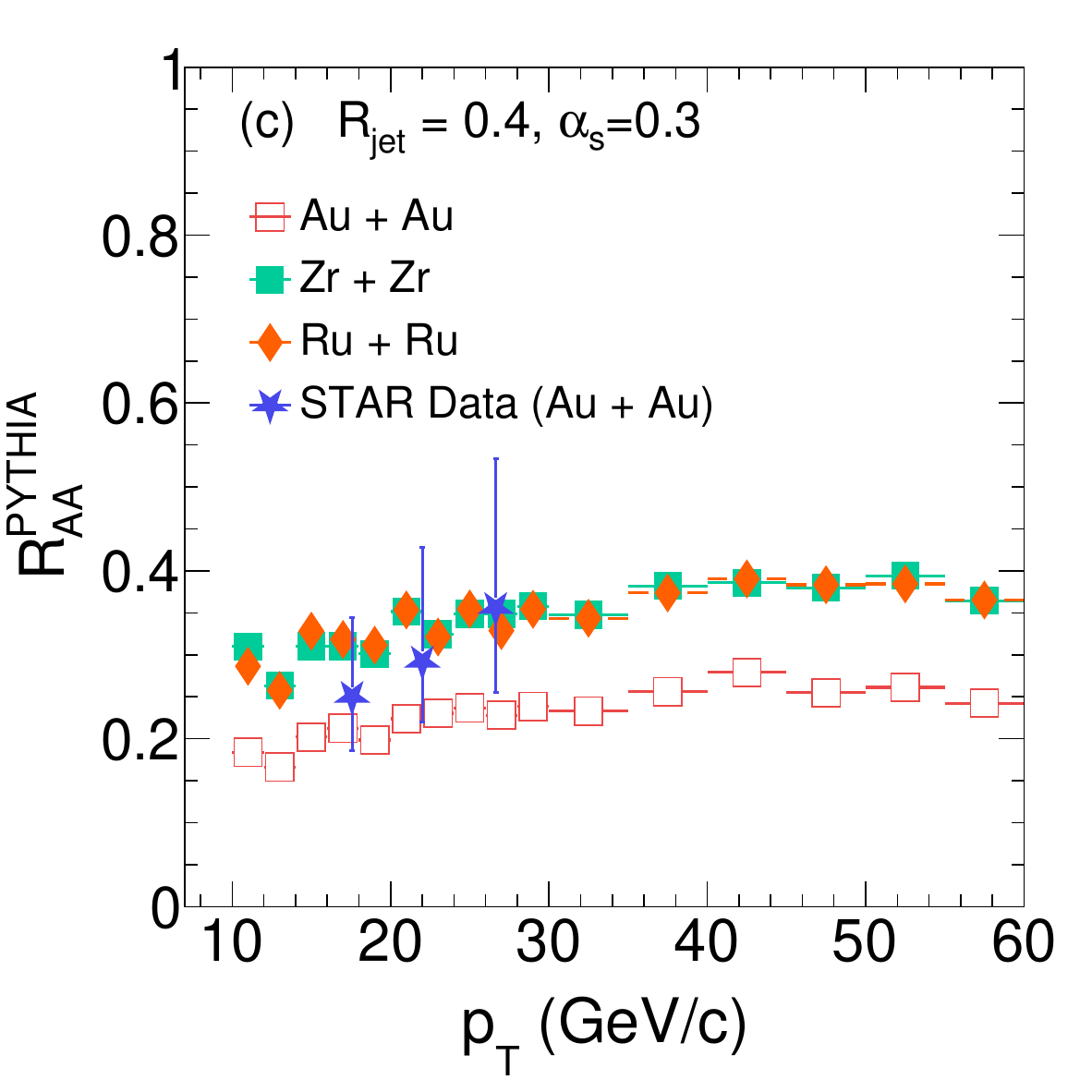}\\
	\caption{(Color online) The jet $R_\mathrm{AA}$ in 0-10\% Au+Au, Zr+Zr and Ru+Ru collisions at $\sqrt{s_\mathrm{NN}}=200$~GeV, compared to the STAR data of Au+Au collisions~\cite{STAR:2020xiv}, upper panel for jet cone size $R=0.2$, middle for $R=0.3$ and lower for $R=0.4$.}
	\label{fig1:RAA}
\end{figure}

Figure~\ref{fig1:RAA} illustrates the nuclear modification factor of inclusive jets as a function of the jet $p_\mathrm{T}$ in central Zr+Zr, Ru+Ru, and Au+Au collisions at $\sqrt{s_\mathrm{NN}}=200$~GeV. Different panels present results with various jet radii. A comparison to the STAR data of Au+Au collisions is also included~\cite{STAR:2020xiv}. The LBT calculation reasonably describes the STAR data in Au+Au collisions, providing a reliable baseline for extending the study to Zr+Zr and Ru+Ru collisions recently conducted at RHIC. It is found that jets still exhibit a significant amount of suppression at Zr+Zr and Ru+Ru collisions, though apparently weaker than that observed in Au+Au collisions. This is due to the smaller size of QGP produced in central Zr+Zr (Ru+Ru) collisions than in central Au+Au collisions. Results in Zr+Zr and Ru+Ru collisions are comparable to each other because of their similar system sizes. Across all systems studied, the jet $R_\mathrm{AA}$ increases with the jet $p_\mathrm{T}$, suggesting a smaller fractional energy loss of the jets at higher $p_\mathrm{T}$. This is partly due to the longer formation time of harder partons that delay the parton-medium interactions and reduce the jet energy loss, as discussed in our earlier work~\cite{Zhang:2022ctd}. However, compared to the experimental data of Au+Au collisions, the rising trend of the jet $R_\mathrm{AA}$ with $p_\mathrm{T}$ appears weaker from our model calculation. This could result from the constant $\alpha_s$ we use, which in principle can decrease for more energetic partons~\cite{JETSCAPE:2022jer}.

\subsection{Jet shape}

We further investigate the energy redistribution within jets across different systems using the jet shape observable. Jet shape quantifies the radial energy distribution relative to the jet axis. The differential jet shape function, denoted as $\rho(r)$, is defined as follows, 
\begin{equation}
\large
\rho(r) = \frac{1}{\delta r}\frac{1}{N_\mathrm{jet} }\sum_\mathrm{jet}\frac{\sum_{\mathrm{track}\in (r-\frac{\delta r}{2}, r+\frac{\delta r}{2})} p_\mathrm{T,track}}{p_\mathrm{T,jet}},
\end{equation}
where $r = \sqrt{\Delta\eta^2+\Delta\phi^2}$ represents the distance between a track (parton) and the jet axis. For each jet, the energy of particles within each annular ring (defined by $\delta r$) is summed and divided by the total jet energy. This value is then averaged across all inclusive jets within a given $p_\mathrm{T}$ range.

\begin{figure}[tbp!]
	\centering
    \vspace{-20pt}
	\includegraphics[scale=0.36]{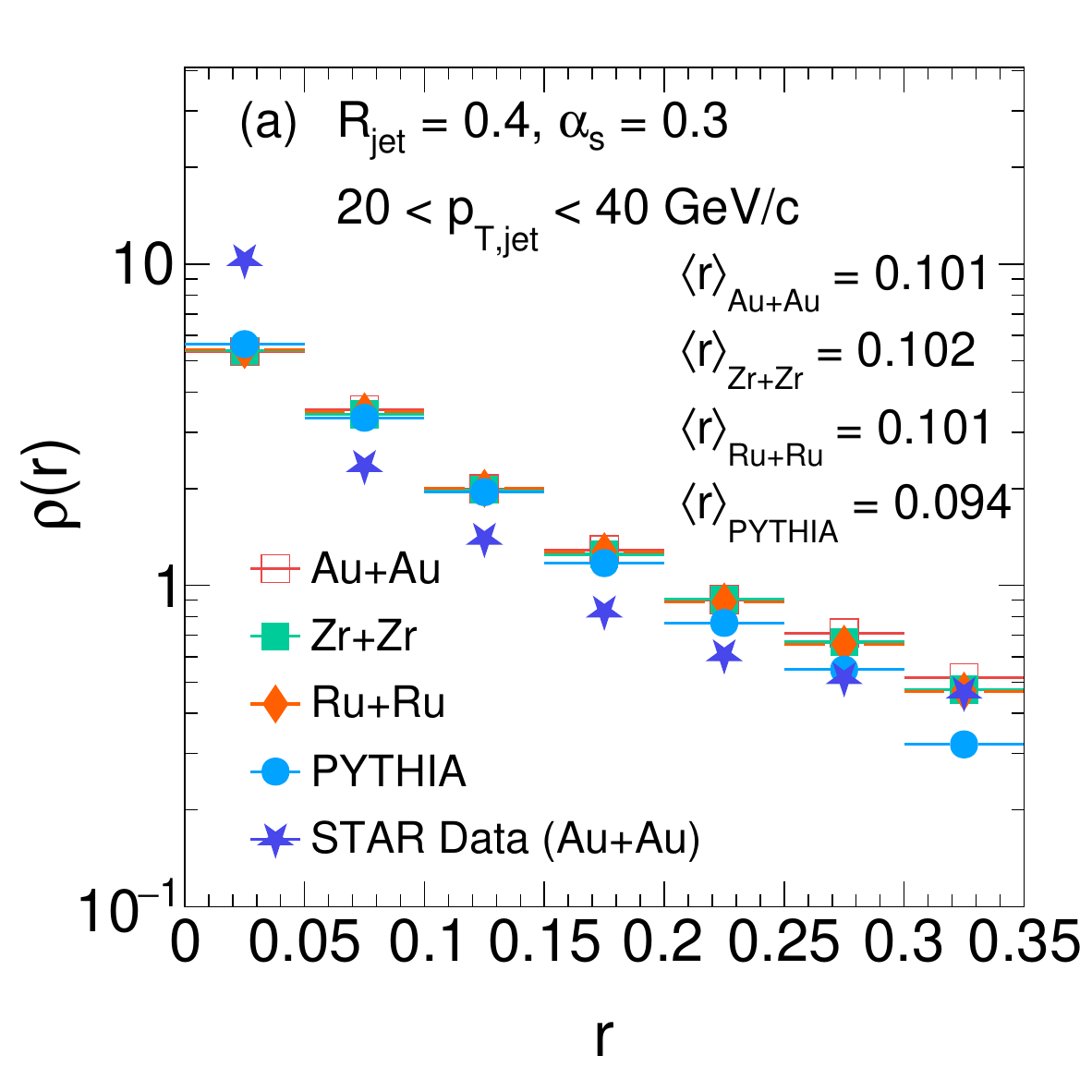}
	\includegraphics[scale=0.36]{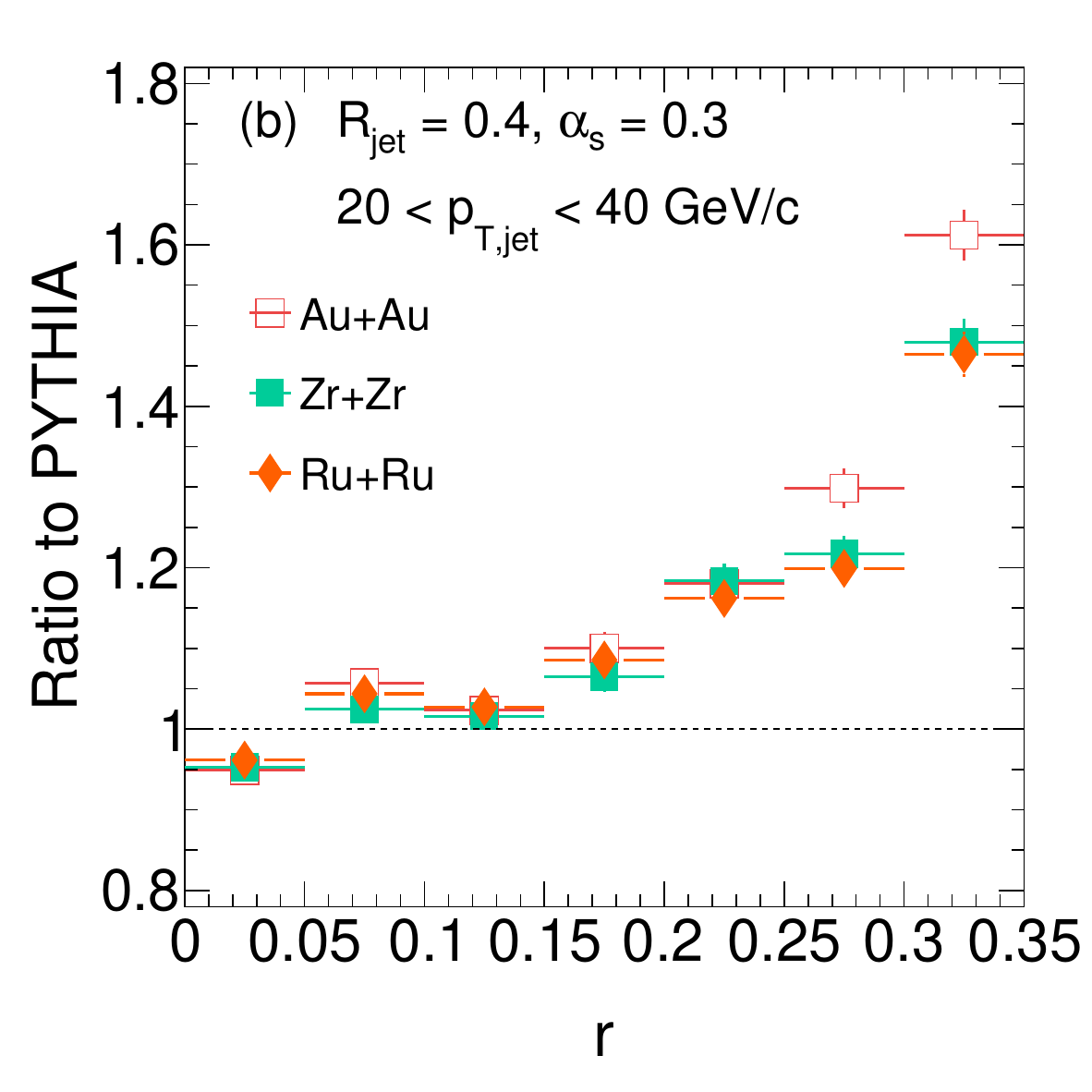}
	\caption{(Color online) (a) Shape of $R=0.4$ jets in 0-10\% Au+Au, Zr+Zr and Ru+Ru collisions simulated by LBT, and in $p$+$p$ collisions simulated by PYTHIA, compared to the STAR data for Au+Au collisions~\cite{Oh:2020yyn}. (b) Ratios of jet shape in heavy-ion collisions to that in $p$+$p$ collisions. }
	\label{fig2:jetshape}
\end{figure}

Shown in Fig.~\ref{fig2:jetshape}~(a) is the shape of $R = 0.4$ jet calculated using partons with $p_\mathrm{T}$ larger than 0.2~GeV/$c$ in central Au+Au, Zr+Zr and Ru+Ru collisions. Results of the three heavy-ion systems from the LBT calculation are compared with that of $p+p$ collisions from the PYTHIA simulation. Significant deviation can be observed between our result and the STAR data in Au+Au collisions~\cite{Oh:2020yyn}, which is largely contributed by the failure of our model calculation on jet shape in $p+p$ collisions, due to both the lack of hadronization process in our present work~\cite{JETSCAPE:2019udz} and the inaccuracy of current Monte-Carlo simulations in literature for vacuum parton showers when jet energy is not sufficiently high. However, this should not prevent us from exploring the sensitivity of the nuclear modification of jet shape on the system size, which can still be tested by the ongoing STAR experiment in Zr+Zr (Ru+Ru) collisions. 
\ approx
In Fig.~\ref{fig2:jetshape} (b), we present the jet shape ratio between A+A collisions modeled by LBT and $p$+$p$ collisions modeled by PYTHIA. In the region of $r\approx 0$, the ratios for all A+A systems are smaller than one due to the energy loss of the leading parton near the jet axis or deflection of the leading parton direction by the medium. Notably, according to a previous study~\cite{Chang:2019sae}, the nuclear modification of jet shape near $r\approx 0$ depends on the jet $p_\mathrm{T}$: small $p_\mathrm{T}$ jets exhibit a ratio below one, while large $p_\mathrm{T}$ jets exhibit a ratio above unity. Furthermore, in that calculation, small $p_\mathrm{T}$ quark jets tend to exhibit an enhancement at $r\approx 0.1$  compared to small $p_\mathrm{T}$ gluon jets. This feature is also seen in our results here for A+A collisions at the RHIC energy, where quark jets dominate.
From the figure, we see the ratio gradually increases with $r$ and significantly surpasses unity at large $r$, indicating the energy inside the jet cone is redistributed to larger angles in heavy-ion collisions compared to PYTHIA. This trend aligns with previous findings from various model calculations~\cite{Park:2018acg, JETSCAPE:2018vyw, CMS:2013lhm, Tachibana:2017syd}. In addition to the $p_\mathrm{T}$ broadening of jet partons induced by the medium, the jet-induced medium excitation, represented by recoil partons in LBT, plays a crucial role in transporting jet energy to wider angles. 

The energy redistribution observed in the jet shape above exhibits a clear system size dependence in the large $r$ region, with Au+Au results showing a greater enhancement than Zr+Zr and Ru+Ru. This discrepancy is attributed to jets in Au+Au collisions experiencing stronger interactions with the larger QGP medium, and therefore transferring more energy from the center to wider angles compared to those in smaller size Zr+Zr and Ru+Ru collisions.

\subsection{Jet mass}

Jet mass is defined as the magnitude of the sum of the four-momenta of all constituents within a jet, expressed as
\begin{equation}
\large
M=\Big|\sum_{i\in \mathrm{jet}} p_i\Big|=\sqrt{E^2-p^2},
\end{equation}
where $p_i$ represents the four-momentum of the $i$-th jet constituent, and $E$ and $p$ denote the total energy and three-momentum of the given jet. In $p$+$p$ collisions, the jet mass quantifies the virtuality of the ancestor parton created from hard collisions that initializes the jet shower. In A+A collisions, the jet mass can be modified by jet-medium interactions.

\begin{figure}[tbp!]
	\centering
    \vspace{-20pt}
	\includegraphics[scale=0.36]{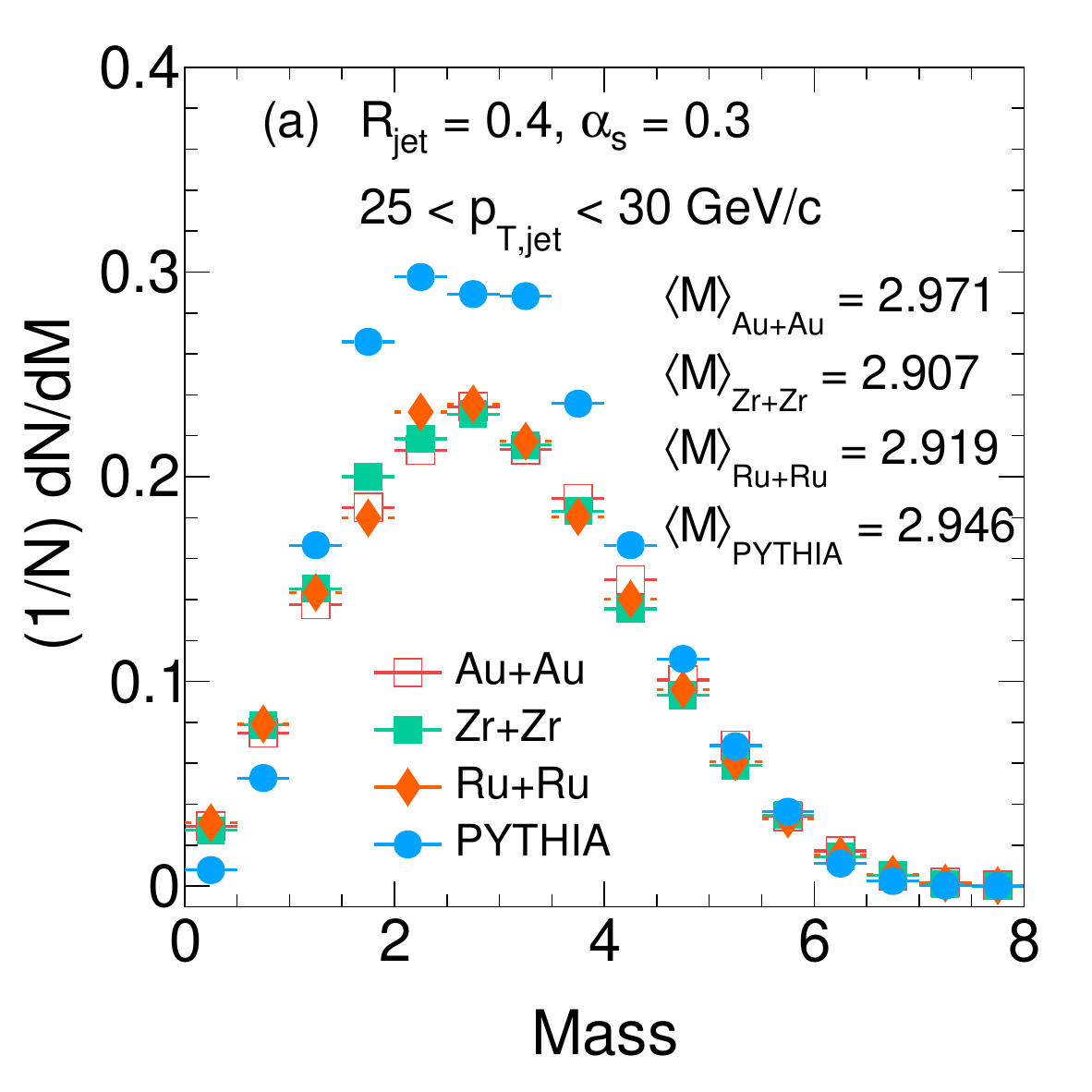}
	\includegraphics[scale=0.36]{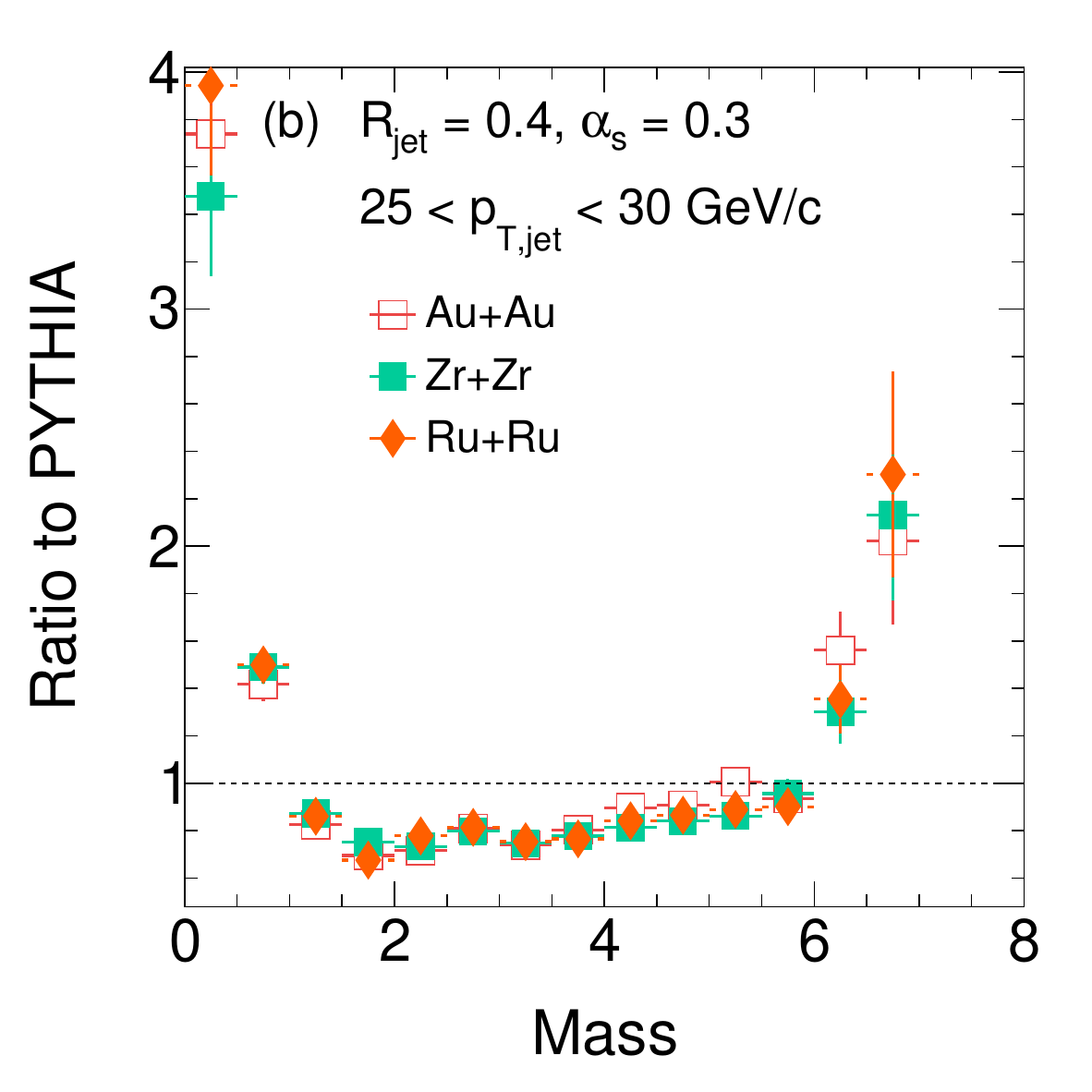}
	\caption{(Color online) (a) Mass of $R=0.4$ jets in 0-10\% Au+Au, Zr+Zr and Ru+Ru collisions simulated by LBT, and in $p$+$p$ collisions simulated by PYTHIA. (b) Ratios of jet mass in heavy-ion collisions to that in $p$+$p$ collisions.}
	\label{fig3:jetmass}
\end{figure}

In Fig.~\ref{fig3:jetmass}~(a), we display the mass distributions for jets with $R=$ 0.4 and 25 $< p_\mathrm{T} < 30$~GeV/$c$ in different collision systems, with the values of average mass listed in the figure. As discussed in Ref.~\cite{STAR:2021lvw}, the hadronization process can shift the mass distribution of jets towards a larger mass region. In Fig.~\ref{fig3:jetmass}~(b), we present the ratios of jet mass between A+A and $p$+$p$ collisions. One can observe enhancement of the mass distribution in A+A collisions relative to $p$+$p$ collisions at both small and large mass regions. We note that the nuclear modification of jet mass is driven by the interplay between several effects: (a) the jet energy loss, or transport of energy outside the jet cone, results in a decrease in mass; (b) selection bias -- jets observed in A+A collisions after suffering energy loss originate from jets with higher $p_\mathrm{T}$ and thus higher mass compared to jets observed in $p$+$p$ collisions with equivalent final $p_\mathrm{T}$; (c) medium response, or recoil partons transferred from medium backgrounds to jets, results in an increase in mass. As shown in Ref.~\cite{Luo:2019lgx}, the enhancement of ratio at the high mass end is mainly driven by medium response within the LBT model. This enhancement at high mass would naturally cause the suppression of the ratio between two normalized distributions at lower mass. Possibly due to the cancellation between effects of jet energy loss and medium response discussed above, no clear difference between Au+Au and Zr+Zr (Ru+Ru) collisions has been observed for jet masses in either their average values or their distributions.  


\subsection{Jet girth}



Jet girth is the $p_\mathrm{T}$-weighted width of a jet, defined as
\begin{equation}
\large
g = \sum_{i\in \mathrm{jet}}\frac{p_{\mathrm{T},i} }{p_\mathrm{T,jet} }r_{\mathrm{jet},i},
\end{equation}
where $p_\mathrm{T,jet}$, $p_{\mathrm{T},i}$ and $r_{\mathrm{jet},i}$ respectively represent the transverse momentum of the jet, the transverse momentum of its $i$-th constituent, and the distance from the $i$-th constituent to the jet axis in the $(\eta,\phi)$ plane. The jet girth provides an additional way to quantify the geometry of a jet. The variation of the jet girth in A+A collisions with respect to $p$+$p$ collisions depends on jet-medium interactions: broadening of jets inside the QGP increases their girths while narrowing of jets decreases their girths.

To calculate the jet girth, we follow the STAR methodology for Au+Au collisions~\cite{starjetshape} to reconstruct jets using the anti-$k_\mathrm{T}$ algorithm with the ``HardCore'' jet selection. The ``HardCore'' selection requires parton $p_\mathrm{T} >2$~GeV/$c$ within $|\eta|<1$. Jets are confined to $|\eta|<1-R$ and must have at least two constituents. The same sample is used for calculating the jet momentum dispersion $p_\mathrm{T}D$ in the next subsection. Under the ``HardCore'' selection, the contribution of soft partons from jet splitting and medium response is suppressed, which implies that the girth and the momentum dispersion calculations focus on the characteristics of a few high-$p_\mathrm{T}$ constituents, i.e. the ``hard cores'' of jets. 

\begin{figure}[tbp!]
	\centering
    \vspace{-20pt}
	\includegraphics[scale=0.36]{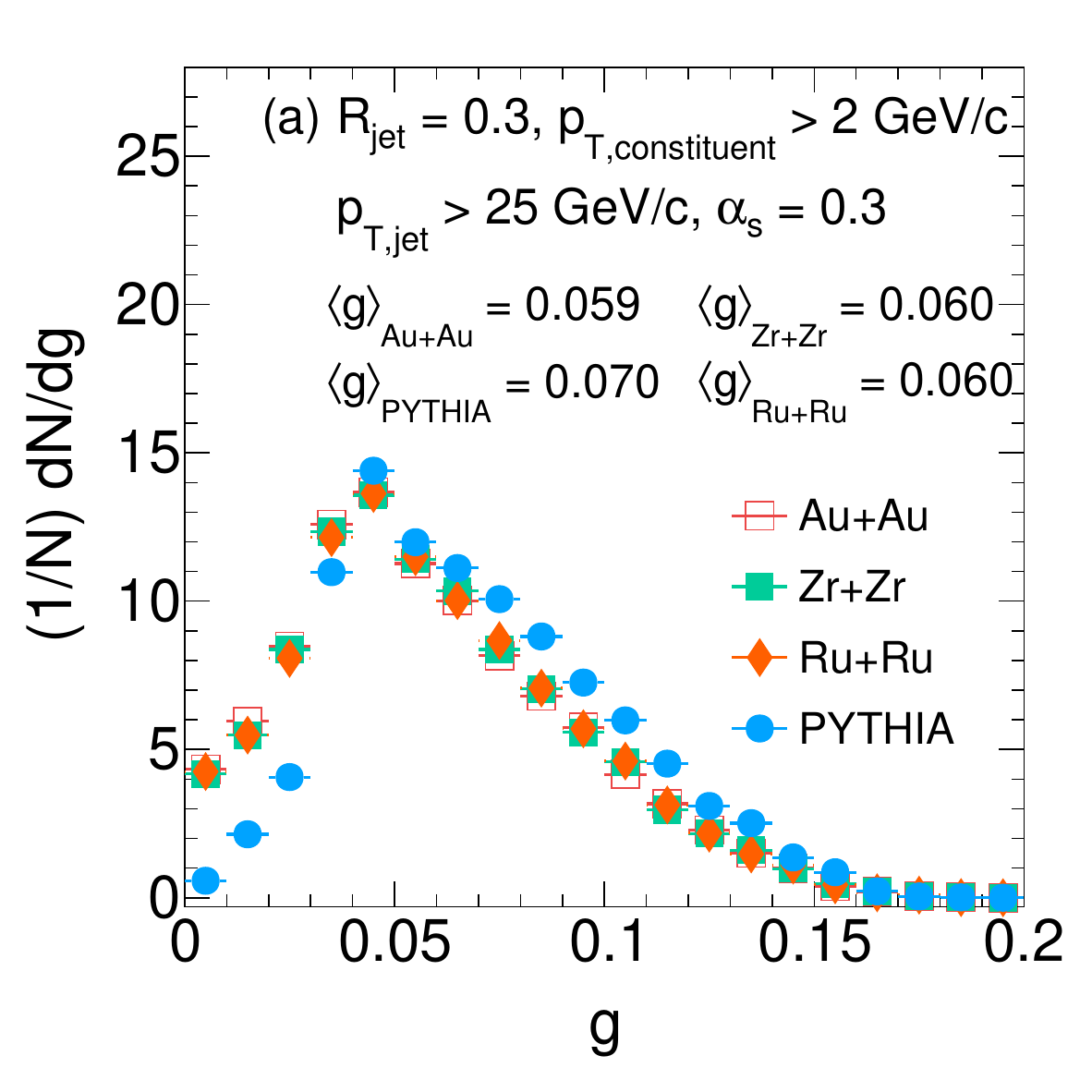}
	\includegraphics[scale=0.36]{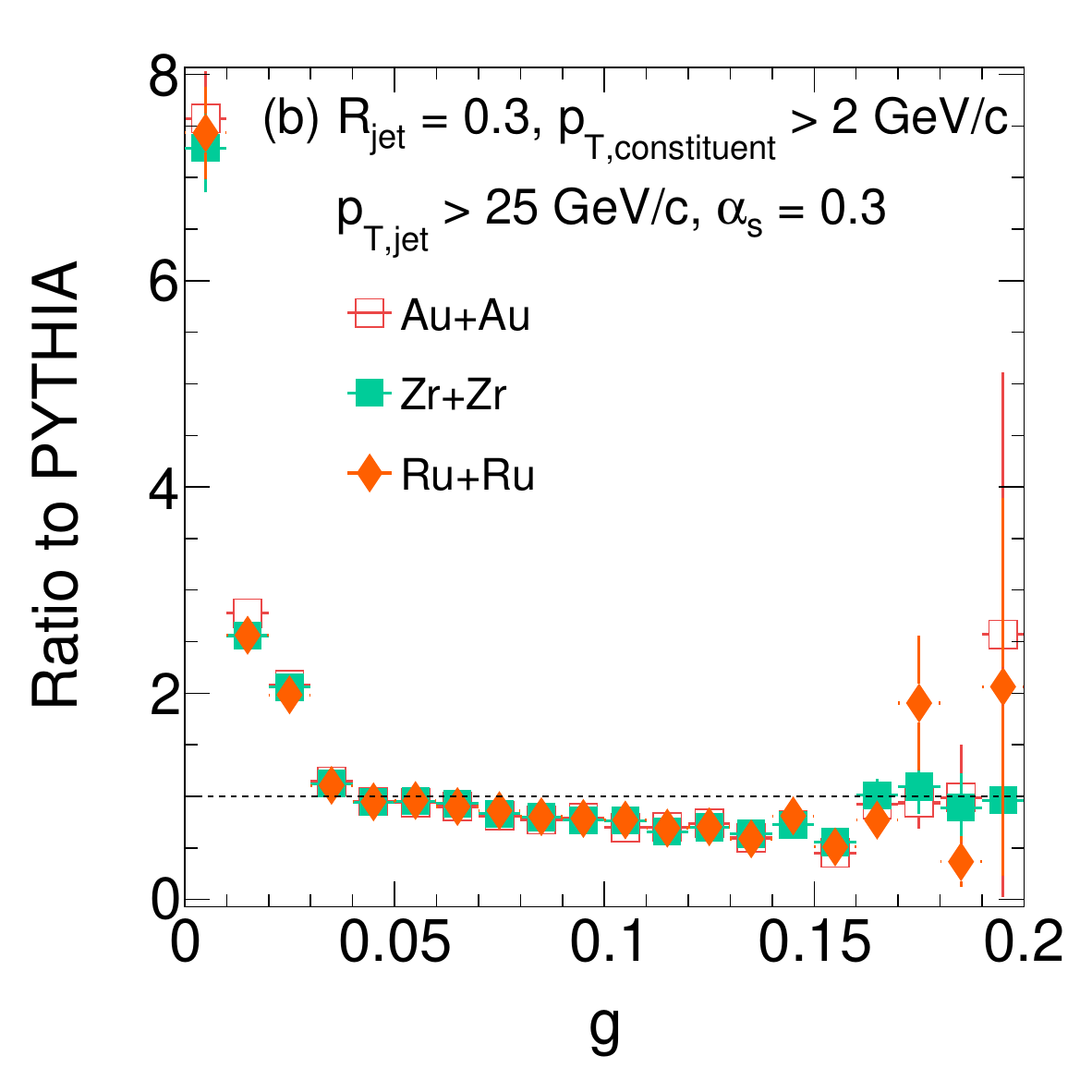}
	\caption{(Color online) (a) Girth of $R=0.3$ jets in 0-10\% Au+Au, Zr+Zr and Ru+Ru collisions simulated by LBT, and in $p$+$p$ collisions simulated by PYTHIA. (b) Ratios of jet girth in heavy-ion collisions to that in $p$+$p$ collisions.}
	\label{fig4:jetgirth}
\end{figure}

Figure~\ref{fig4:jetgirth}~(a) displays the girth distribution of $R=0.3$ jets for different collision systems, with the average values of girth summarized in the plot. Figure~\ref{fig4:jetgirth}~(b) shows the corresponding ratios between A+A and $p$+$p$ collisions. One can clearly observe that the girth distributions in heavy-ion systems shift towards lower girth region compared to $p$+$p$ collisions, indicating the hard cores of jets become narrower in the QGP medium. This shift is likely attributed to the selection bias. With a given range of the final $p_\mathrm{T}$, compared to jets in $p$+$p$ collisions, jets observed in A+A collisions originate from higher $p_\mathrm{T}$ ancestor partons that shower into narrower jets. In addition, compared to harder partons at smaller angles, softer partons at wider angles inside jets are easier to be suppressed by the medium and thus fail the ``HardCore" selection. As a result, jets in A+A collisions that survive the ``HardCore" selection appear narrower than their companions in $p+p$ collisions.
However, when comparing Zr+Zr and Ru+Ru to Au+Au collisions, the jet girth does not exhibit a clear dependence on the system size.

\subsection{Jet momentum dispersion}

Jet momentum dispersion $p_\mathrm{T}D$ characterizes the spread of particle momenta within a jet. 
It is defined as the second moment of the transverse momentum distribution of jet constitutes as follows,
\begin{equation}
\large
p_\mathrm{T}D=\frac{\sqrt{ \sum_{i\in \mathrm{jet}}p^2_{\mathrm{T},i} } }{\sum_{i\in \mathrm{jet}}p_{\mathrm{T},i}}.
\end{equation}
A jet with low $p_\mathrm{T}D$ indicates the particles within the jet have similar momenta, while a jet with high $p_\mathrm{T}D$ implies a broad range of particle momenta within the jet.
For instance, in an extreme scenario where a single parton carries the entire jet momentum, $p_\mathrm{T}D$ is 1. Conversely, if there are numerous soft constituents with low momenta, $p_\mathrm{T}D$ is close to 0. Therefore, this $p_\mathrm{T}D$ reflects the dynamics of both parton shower and hadronization processes. Since we have not included hadronization in the present study, our results here are not expected to be quantitatively compared to experimental data.

\begin{figure}[tbp!]
	\centering
    \vspace{-20pt}
	\includegraphics[scale=0.36]{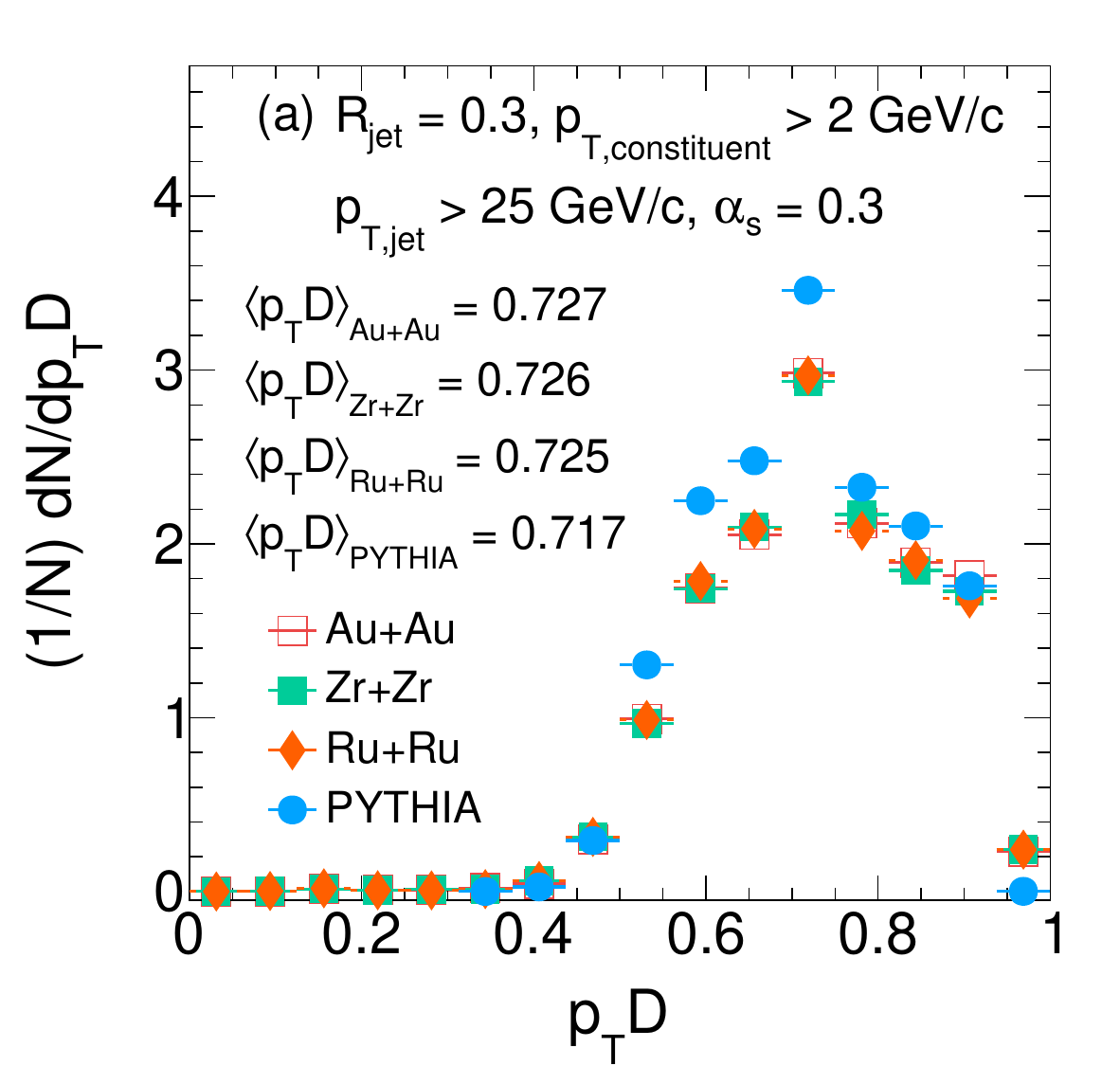}
	\includegraphics[scale=0.36]{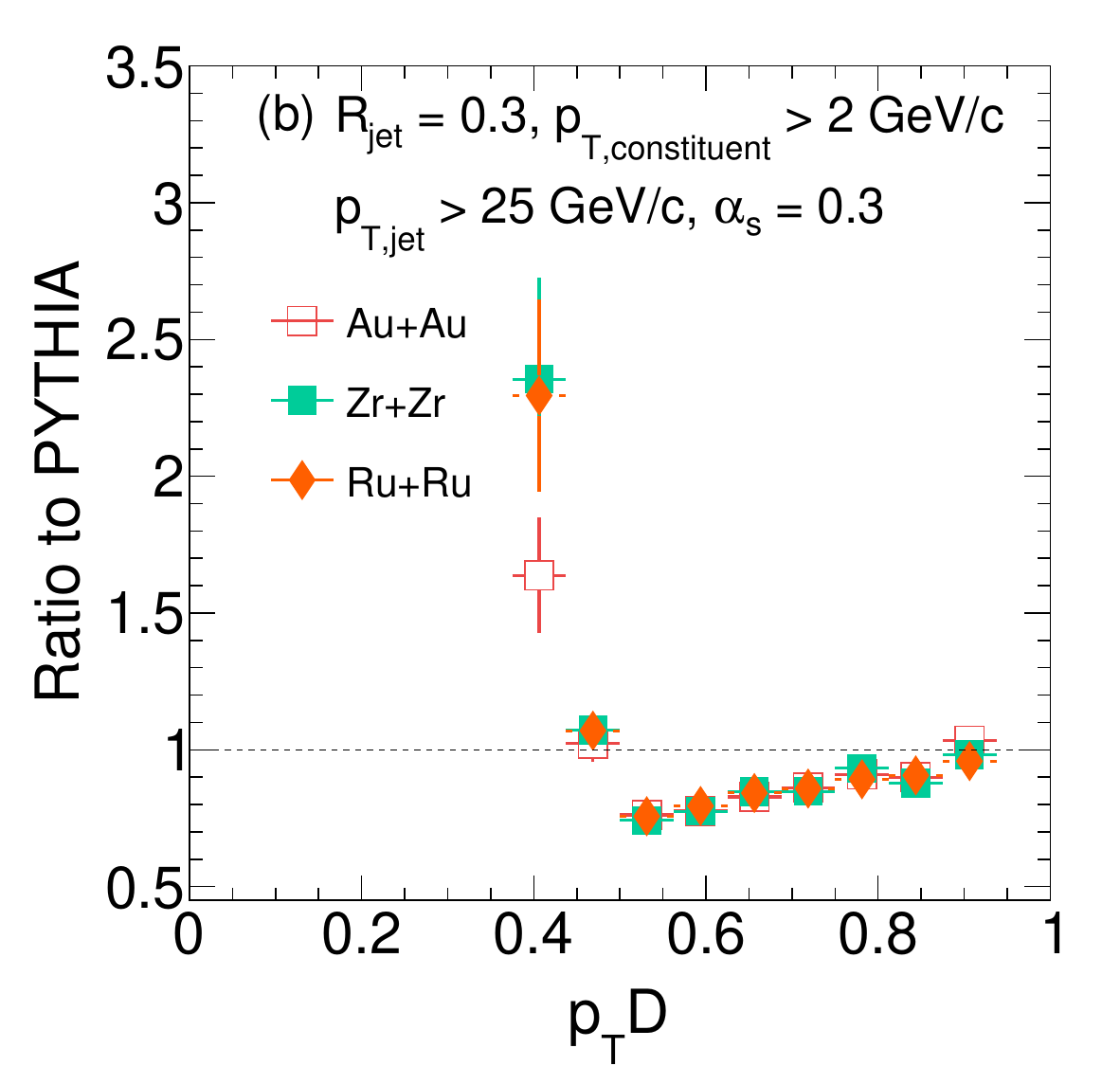}
	\caption{(Color online) (a) Momentum dispersion of $R=0.3$ jets in 0-10\% Au+Au, Zr+Zr and Ru+Ru collisions simulated by LBT, and in $p$+$p$ collisions simulated by PYTHIA. (b) Ratios of jet momentum dispersion in heavy-ion collisions to that in $p$+$p$ collisions.}
	\label{fig5:jetptd}
\end{figure}


Shown in Fig.~\ref{fig5:jetptd}~(a) are the distributions of jet $p_\mathrm{T}D$ in different collision systems, with their average values listed in the plot. The ratios between A+A and $p$+$p$ collisions are presented in Fig.~\ref{fig5:jetptd}~(b). Due to the ``HardCore" selection implemented here, few jet events are found below $p_\mathrm{T}D\approx 0.3$. One can observe an enhancement of the $p_\mathrm{T}D$ distributions in A+A collisions relative to $p$+$p$ collisions below $p_\mathrm{T}D\approx 0.45$, which can be caused by medium-induced jet splittings. Due to the self-normalization of the $p_\mathrm{T}D$ distributions in A+A and $p+p$ collisions, enhancement of their ratio at low $p_\mathrm{T}D$ causes suppression at high $p_\mathrm{T}D$. The increasing trend of the ratio above $p_\mathrm{T}D \approx 0.5$ indicates
jets with limited number of hard constituents are less modified by the medium and therefore are easier to pass the ``HardCore" selection than jets comprising multiple soft constituents. 
When comparing Zr+Zr and Ru+Ru to Au+Au collisions, the jet momentum dispersion shows a weak dependence on the system size.


\section{Conclusion}
\label{sec:summary}

We have investigated the dependence of various jet observables on the system size of relativistic heavy-ion collisions, aiming at pinpointing the size-sensitive jet observables and offering predictions for the upcoming jet measurements in Zr+Zr and Ru+Ru collisions at RHIC. By employing the LBT model to simulate the interactions between jet partons and  color deconfined QCD media and implementing consistent jet analysis schemes utilized in STAR experiments, we have found jet observables, including the nuclear modification factor $R_\mathrm{AA}$ of jets, jet shape $\rho$, jet mass $M$, jet girth $g$, and jet momentum dispersion $p_\mathrm{T}D$ are significantly modified in Au+Au, Zr+Zr and Ru+Ru collisions relative to $p+p$ collisions at $\sqrt{s_\mathrm{NN}}=200$~GeV, but with different sensitivities to the system size of heavy-ion collisions. 

We have observed a less suppressed $R_\mathrm{AA}$ of inclusive jets in smaller size Zr+Zr and Ru+Ru collisions compared to larger size Au+Au collisions, indicating a clear medium size effect on jet quenching. Furthermore, we have observed a stronger enhancement of the jet shape at large angles with increasing system size, suggesting more energy transport to large angles when jets scatter through a larger QGP medium. 
If confirmed, this observation can further improve our understanding on how energy is redistributed inside jets, specifically elucidating the role of the medium response in various medium environments. In contrast, although jet mass, girth and momentum dispersion also unveil strong QGP effect in A+A collisions, no significant differences in these observables have been observed between Zr+Zr (Ru+Ru) and Au+Au collisions at the RHIC energy. This can be caused by the cancellation between different effects on the internal structures of jets, such as jet energy loss, momentum broadening, medium response and selection bias. Therefore, one should select appropriate observables for jet measurements in Zr+Zr (Ru+Ru) collisions at RHIC for the purpose of investigating the system size dependence of jet-medium interactions. 



Although our current calculation is at the partonic level, it offers valuable insights into the sensitivities of different jet observables to the system size of heavy-ion collisions, and provides timely guidance for anticipated upcoming experimental measurements at RHIC that progress towards bridging the jet quenching phenomena between large and small collision systems. In order to achieve more precise quantitative predictions for experimental measurements, we will include a hadronization scheme that can be simultaneously applied to $p+p$ and A+A collisions in our future work. The $p+p$ calculation for jet substructures will also be improved by either re-tuning Pythia once the corresponding data at RHIC are available, or exploring event generators that implement NLO contributions to parton showers, e.g., Sherpa~\cite{Sherpa:2019gpd}. A reliable $p+p$ baseline for jet observables is crucial for our further study on their nuclear modification in A+A collisions. The system size dependence of these jet observables can also help improve our understanding of the dynamics of jet-QGP interactions, such as the path length dependence of parton energy loss, momentum broadening, and medium response.

\begin{acknowledgments} 
We thank Rongrong Ma and Isaac Mooney for helpful discussions. This work is supported by the National Natural Science Foundation of China (NSFC) under Grant Nos.~12175122, 2021-867, 11890710, 11890713, 12105156, 14-547, Shandong Provincial Natural Science Foundation project ZR2021QA084,  and National Key Research and Development Program of China under Contract No. 2022YFA1604900. 
\end{acknowledgments}


\bibliographystyle{h-physrev5}
\bibliography{ref_0813}


\end{document}